\documentclass{article}

\usepackage[preprint,nonatbib]{neurips_2024}

\usepackage[utf8]{inputenc} 
\usepackage[T1]{fontenc}    
\usepackage{url}            
\usepackage{booktabs}       
\usepackage{amsfonts}       
\usepackage{nicefrac}       
\usepackage{microtype}      

\usepackage{color}
\definecolor{lightblue}{RGB}{0, 110, 204}
\usepackage{amsfonts,amssymb}
\usepackage{amsmath,bm}
\usepackage[table]{xcolor}
\usepackage{graphicx}
\usepackage{algorithmic}
\usepackage{algorithm}
\usepackage{pifont}

\usepackage{amsthm}
\usepackage[pagebackref=true,breaklinks=true,letterpaper=true,citecolor=lightblue,colorlinks,bookmarks=false]{hyperref}
\usepackage{bbding}
\usepackage{wrapfig}
\usepackage{subfig}

\def\eg{\textit{e.g.}}
\def\ie{\textit{i.e.}}

\newcommand{\para}[1]{\vspace{.05in}\noindent\textbf{#1}}

\title{Generative Enhancement for 3D Medical Images}

\author{
Lingting Zhu$^1$,
Noel Codella$^2$\footnotemark[1],
Dongdong Chen$^2$\footnotemark[1],
Zhenchao Jin$^1$,
Lu Yuan$^2$,
Lequan Yu$^1$\footnotemark[2]\\
$^1$The University of Hong Kong,
$^2$Microsoft \\
\texttt{ltzhu99@connect.hku.hk, lqyu@hku.hk}
}

\begin{document}

\renewcommand{\thefootnote}{\fnsymbol{footnote}} 
\footnotetext[1]{Equal contribution project leads.}
\footnotetext[2]{Corresponding author.}
\renewcommand{\thefootnote}{\arabic{footnote}}

\maketitle

\begin{abstract}
The limited availability of 3D medical image datasets, due to privacy concerns and high collection or annotation costs, poses significant challenges in the field of medical imaging. 
While a promising alternative is the use of synthesized medical data, there are few solutions for realistic 3D medical image synthesis due to difficulties in backbone design and fewer 3D training samples compared to 2D counterparts.
In this paper, we propose \textbf{GEM-3D}, a novel generative approach to the synthesis of 3D medical images and the enhancement of existing datasets using conditional diffusion models. 
Our method begins with a 2D slice, noted as the informed slice to serve the patient prior, and propagates the generation process using a 3D segmentation mask. 
By decomposing the 3D medical images into masks and patient prior information, GEM-3D offers a flexible yet effective solution for generating versatile 3D images from existing datasets. 
GEM-3D can enable dataset enhancement by combining informed slice selection and generation at random positions, along with editable mask volumes to introduce large variations in diffusion sampling.
Moreover, as the informed slice contains patient-wise information, GEM-3D can also facilitate counterfactual image synthesis and dataset-level de-enhancement with desired control. 
Experiments on brain MRI and abdomen CT images demonstrate that GEM-3D is capable of synthesizing high-quality 3D medical images with volumetric consistency, offering a straightforward solution for dataset enhancement during inference. The code is available at {\href{https://github.com/HKU-MedAI/GEM-3D}{https://github.com/HKU-MedAI/GEM-3D}}.

\end{abstract}

\section{Introduction}

In the realm of medical image analysis, 3D medical images acquired from computed tomography (CT), magnetic resonance imaging (MRI), and ultrasound imaging, serve a variety of purposes, including diagnostics, medical education, surgical planning, and patient communication \cite{duncan2000medical, singh20203d}. 
Nonetheless, the existence of privacy concerns within the medical sector obstructs the development of extensive datasets, thereby hampering advancements in several downstream tasks.
Furthermore, the collection and annotation of 3D medical images are costly due to the expenses associated with scanning and the need for expert-level labor. 
Consequently, a considerable disparity in dataset size exists when compared to other rapidly emerging domains, such as text-to-video generation. For example, EMU VIDEO~\cite{girdhar2023emu} is trained on a dataset of 34 million licensed video-text pairs, whereas publicly accessible 3D medical image datasets typically consist of a mere several dozen to a few hundred 3D volumes~\cite{antonelli2022medical}.

The advent of generative models \cite{goodfellow2014generative, sohl2015deep,ho2020denoising} has led to significant advancements in data generation and downstream tasks across various domains \cite{isola2017image, sankaranarayanan2018learning, azizi2023synthetic, ramesh2021zero, poole2022dreamfusion, wu2023diffumask, rombach2022high}. 
However, realistic 3D medical image synthesis remains a formidable challenge: First, the generation of 3D medical images presents difficulties in designing 3D backbone architectures, as traditional 3D backbones require substantial memory resources. Second, the relatively small size of 3D medical image datasets results in training difficulty and poor generalization of the model. 
A viable strategy to address these challenges involves treating 3D medical volumes as a sequence of slices and then employing memory-efficient diffusion frameworks based on 2D or pseudo-3D diffusion models~\cite{peng2023generating, han2023medgen3d, zhu2023make}. 
These approaches offer significant advantages in terms of memory-efficient training and training data-efficient usage, facilitating high-fidelity image generation. 
Our work follows the principle design of these works to create a 3D medical image generation framework that relies on volume diffusion, which jointly captures anatomical information within a window of slices and then propagates the conditional generation to form complete 3D volumes.

In this paper, we focus on the topic of Generative Enhancement for 3D Medical Images, addressing the question of how to generate 3D data samples from existing 3D medical image datasets using generative models. This problem comprises two main aspects. First, a high-quality 3D generation capability is required for 3D medical images. Second, the generative models should incorporate condition decoupling, enabling re-sampling from the data and enhancing distribution coverage, even when only given the training dataset. To address these challenges, we present GEM-3D, a novel approach that leverages conditional diffusion models to synthesize realistic 3D medical images and enhance existing datasets.
Specifically, our method enables the synthesis of new medical image volumes from 3D structure masks (\eg, annotated segmentation masks) with optional variations. 
Distinct from the previous works in 3D medical image generation using diffusion models \cite{peng2023generating, han2023medgen3d, zhu2023make, khader2022medical, kim2022diffusion}, a key innovation of our approach involves the decomposition of segmentation mask and patient-prior information in the generation process, which not only significantly improves quality of the generated 3D images but also enables enhancement of existing medical datasets.
Particularly, we achieve this by introducing \textit{informed slice}, which is a 2D slice in 3D volumes and contains prior information on the patient indicative of anatomical appearance, physical position, and scanning patterns, thereby facilitating one-to-many ill-posed mappings within the mask-driven generation. 
Moreover, the incorporation of informed slice enables the combination of existing masks and patient prior information to create counterfactual medical volumes for specific patients or pathological conditions (\eg, particular tumors). 

With the proposed generative process for 3D medical images, we provide a practical solution for dataset enhancement by training the generative model on the original dataset and then performing data re-sampling to improve the distribution coverage of observed information. 
Moreover, by selectively executing mask-driven generation with informed slices from different patients, GEM-3D can facilitate the synthesis of 3D medical images containing patient-specific information, such as anatomical appearance and scanning patterns, thereby creating counterfactual volumes of a specific patient with controllable masks.  
Another potential downstream application of our method is de-enhancement as dataset-level normalization. 
By conducting conditional generation with the same informed slice, GEM-3D offers an optional normalization technique for handling medical images acquired using different protocols across multiple scanners~\cite{bashyam2022deep}. 

In summary, our contributions are as follows:
\begin{itemize}
    \item We propose GEM-3D, a novel 3D medical image generation scheme that integrates patient-specific informed slices and structure mask volumes to synthesize realistic 3D medical images based on conditional volume diffusion.
    \item Our approach not only achieves high-quality generation of 3D medical images but also allows for the creation of new data samples, even when only given the training datasets.
    \item We demonstrate the applicability of GEM-3D for counterfactual image generation with optional control over the informed slices and editable mask volumes and generative de-enhancement in reverse at the dataset level for 3D medical images.
\end{itemize}
\section{Related Works}

\para{Diffusion Models.} In recent years, diffusion models \cite{sohl2015deep, ho2020denoising, song2020denoising} have been extensively studied due to their high fidelity and stable training \cite{dhariwal2021diffusion}, outperforming counterparts such as GANs \cite{goodfellow2014generative} and VAEs \cite{kingma2013auto}. Variants of diffusion models generate samples by gradually denoising from initial Gaussian noises and are trained using a stationary training objective expressed as a reweighted variational lower-bound \cite{ho2020denoising}. Despite their success in achieving impressive results in image generation \cite{rombach2022high, ramesh2022hierarchical, saharia2022photorealistic}, diffusion models have also been applied to other generative tasks, yielding state-of-the-art performances in areas such as text-to-3D \cite{poole2022dreamfusion, lin2023magic3d} and text-to-video \cite{singer2022make, blattmann2023align, girdhar2023emu}.
The latent diffusion model, when combined with a pre-trained variational autoencoder, enables training on limited computational resources while retaining quality and flexibility \cite{rombach2022high}. This model serves as a basic architecture for various generative tasks and facilitates the development of Stable Diffusion models, which have been used as foundation models for tasks such as open-vocabulary segmentation \cite{xu2023open}, semantic correspondence \cite{hedlin2024unsupervised}, and personalized image generation \cite{ruiz2023dreambooth}. 
Controllable generation \cite{mirza2014conditional, isola2017image} is an important topic within the literature of diffusion models that aims to enable models to accept more user controls. There are works achieving multi-condition generation by either training diffusion models from scratch \cite{bao2023one, huang2023composer} or fine-tuning lightweight adapters \cite{zhang2023adding, mou2023t2i, zhao2024uni}.

\para{3D Medical Image Synthesis.} Given the critical importance of data privacy in medical imaging, synthesizing realistic medical images is a promising direction offering potential solutions for numerous applications. Existing research has utilized GANs and diffusion models to achieve satisfactory unconditional generation of medical images and multi-modality MRI in 2D approaches \cite{han2018gan, jiang2023cola, yi2019generative}. Some works have also explored generating images from text prompts. For instance, RoentGen \cite{chambon2022roentgen} fine-tunes Stable Diffusion to synthesize chest X-ray images from radiology reports, while BiomedJourney \cite{gu2023biomedjourney} follows the pipeline of InstructPix2Pix \cite{brooks2023instructpix2pix} to produce counterfactual chest X-ray images with disease progression descriptions. However, 3D medical images are of greater importance as they align with real-world scanning in hospitals. Previous works \cite{kwon2019generation, khader2022medical} combine VAEs and leverage 3D architectures, such as 3D GAN and 3D Diffusion, for volume generation. Nonetheless, these approaches are still limited by the computational demands of 3D architectures. To efficiently generate 3D images, \cite{peng2023generating} employs self-conditional generation autoregressively for generating Brain MRIs. Moreover, MedGen3D \cite{han2023medgen3d} and Make-A-Volume \cite{zhu2023make} establish 3D image generators primarily on 2D or pseudo-3D architectures, mitigating volumetric inconsistency through volumetric refiners or tuning. In this paper, we follow the paradigms of efficient 3D generation in \cite{peng2023generating, han2023medgen3d, zhu2023make} with novel designs. Specifically, we synthesize small volume windows and then propagate the generation in two directions to ultimately form 3D volumes.

\begin{figure*}[t]
\centering
\includegraphics[width=1.0\textwidth]{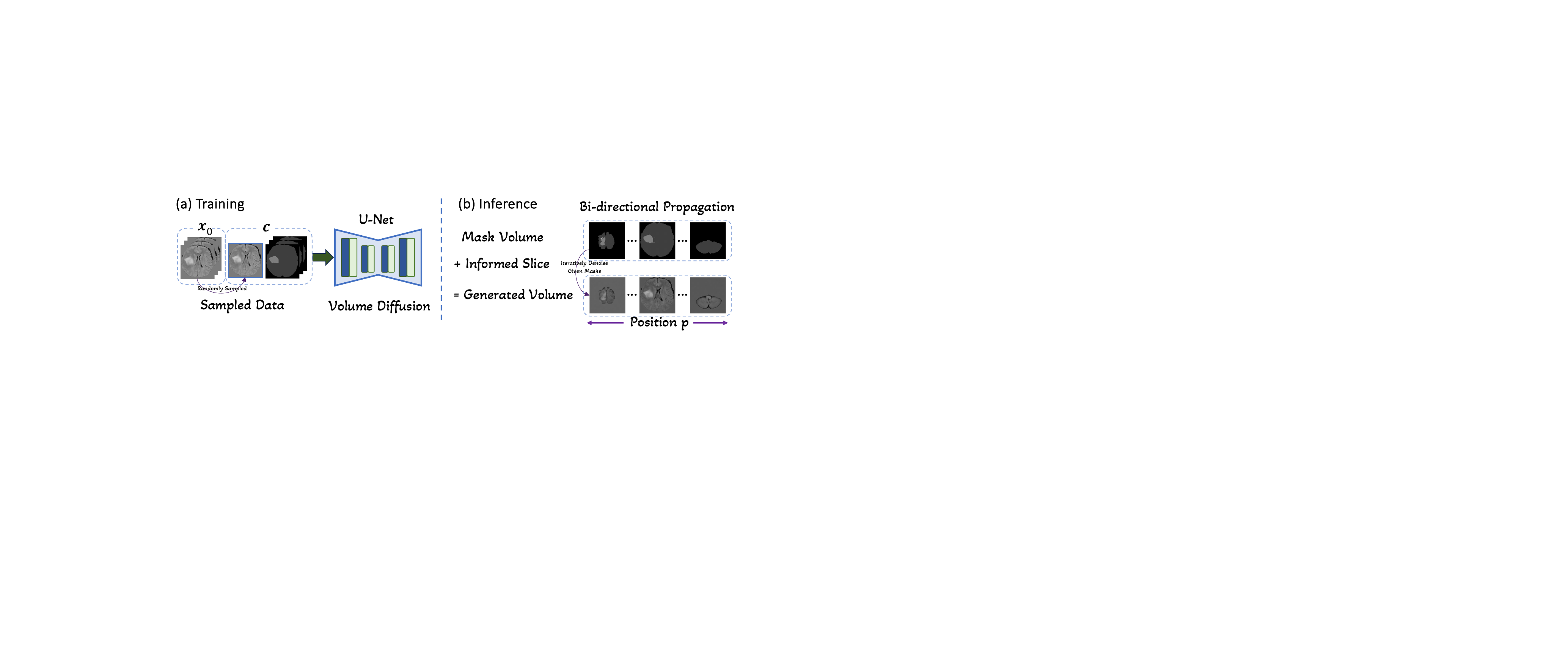}
\caption{\textbf{Overview of GEM-3D Framework.} \textbf{(a)} During training, our method is built upon volume diffusion and we sample a window of images and their corresponding masks as training samples. The informed slice, sampled from the volume window, combines with mask data as conditions. \textbf{(b)} For inference, we decouple conditional generation into the combination of mask volume and informed slice. Our designs include random starting positions, editable mask volume, and a selective or generative informed slice for increased variations, employing bi-directional propagation in sampling.} 
\label{framework}
\vspace{-12pt}
\end{figure*}

\para{Synthetic Data for Downstream Tasks.} 
Synthetic datasets demonstrate the potential to boost downstream tasks \cite{zhang2021datasetgan, li2022bigdatasetgan}. In the context of diffusion models, relevant research includes but not limited to image classification \cite{he2022synthetic, azizi2023synthetic}, semantic segmentation \cite{wang2022semantic, wu2023diffumask, yang2024freemask}, and instance segmentation \cite{zhao2023x}. Studies have shown that synthetic datasets can enhance domain adaptation \cite{sankaranarayanan2018learning} and increase robustness for domain generalization \cite{wu2024datasetdm}. In the medical domain, previous work \cite{han2019synthesizing, zhang2023self, hu2023label, feng2023cheap, chen2024towards} has focused on synthesizing tumors and nodules for segmentation and detection tasks.

\section{Methods}

\para{Overview.} Fig. \ref{framework} provides an overview of the proposed \textbf{GEM-3D} framework, which is designed to synthesize 3D medical images and enhance existing datasets. Given a dataset of 3D medical images $\{(\bm V_i, \bm {M}_i )\}_{i=1}^{N}$, where $\bm V_i \in \mathbb{R}^{H \times W \times Z}$ denotes the $i$-th medical volume sample and $\bm M_i \in \mathbb{R}^{H \times W \times Z}$ represents mask volume, our objective is to train generative models capable of synthesizing high-quality 3D medical images and enhancing the datasets with counterfactual samples.
In this section, we first present the foundation of diffusion based volume synthesis (Section~\ref{sec:3.2}). Next, we discuss informed slice conditioned generation for mask-driven 3D medical image synthesis (Section~\ref{sec:3.3}). To promote generative enhancement, we introduce methodologies that incorporate variations during the sampling stage (Section~\ref{sec:3.4}).

\subsection{Preliminary: Volume Diffusion}
\label{sec:3.2}
Diffusion models \cite{sohl2015deep, ho2020denoising, song2020denoising} aim to learn the reverse process and denoise data samples from noise to clean samples $\bm x_0$ following true data distribution $\bm x_0 \sim q(\bm x)$. In the reverse process, the denoising network $\bm \epsilon_{\bm \theta}(\cdot)$ predicts the noise of corrupted data $\bm x_t$ at timestep $t \in \{1,2,...,T\}$ and constructs the parameterized Gaussian transition $p_\theta(\bm x_{t-1} | \bm x_t)= \mathcal{N}(\bm x_{t-1}; \bm \mu_{\bm \theta}(\bm x_t, t), \sigma^2_t\bm I)$. The denoising network is typically built upon U-Net \cite{ronneberger2015u} and trained with mean squared error \cite{ho2020denoising}:
\begin{align}
L(\bm \theta) = \mathbb{E}_{\bm x_0, \bm c, t, \bm \epsilon \sim \mathcal{N}(0, 1)}\left[\left\| \bm \epsilon - \bm \epsilon_{\bm \theta}(\sqrt{\bar\alpha_t} \bm x_0 + \sqrt{1-\bar\alpha_t} \bm \epsilon, t, \bm c) \right\|^2\right],
\label{eq:losssimple}
\end{align}
where $\sqrt{\bar\alpha_t}$ and $\sqrt{1-\bar\alpha_t}$ control the noise schedule, and $\bm c$ represents the conditional information in general conditional generation settings, jointly sampled with $\bm x_0, \bm c \sim q(\bm x, \bm c)$.

To effectively conduct training and inference, GEM-3D framework adopts the two-stage volumetric diffusion from Make-A-Volume \cite{zhu2023make}. In the training phase, we first train the slice model and then fine-tune volumetric layers. During the inference stage, we autoregressively complete 3D generation based on the conditional generation of small volume windows. The Latent Diffusion Model (LDM) \cite{rombach2022high} serves as the basic structure for slice and volume diffusion, wherein the pre-trained VAE \cite{esser2021taming, kingma2013auto} encodes the slices to latents via $\bm z = \mathcal{E}(\hat{\bm x})$ and reconstructs data from generated latents via $\hat{\bm x} = \mathcal{D}(\hat{\bm z})$.

In the second stage of training, \ie, volumetric tuning, we incorporate pseudo 3D convolutional and attention layers \cite{chollet2017xception, singer2022make} to enhance consistency in the temporal or volumetric dimension \cite{singer2022make, blattmann2023align, zhu2023make}. We treat the volume, consisting of $n$ slices, as an element in this stage and tune the volumetric model with the same timestep across different slices in each volume. Specifically, let $b_v$ and $b_s = b_v n$ represent the volume batch and slice batch, respectively. Our goal is to tune the pseudo 3D layers that operate on the latent feature of the volume batch $\bm f \in \mathbb{R}^{(b_v\times n) \times c\times h\times w}$, outputting $\bm f^{\prime}$ as follows:
\begin{align} \label{lv} \notag
&\bm f^{\prime} \leftarrow \texttt{Rearrange}(\bm f, (b_v\times n)\ c\ h\ w \rightarrow (b_v\times h\times w)\ c\ n),\\ \notag
&\bm f^{\prime} \leftarrow l_v(\bm f^{\prime}),\\ \notag
&\bm f^{\prime} \leftarrow \texttt{Rearrange}(\bm f^{\prime}, (b_v\times h\times w)\ c\ n \rightarrow (b_v\times n)\ c\ h\ w).
\end{align} 
Here, $\texttt{Rearrange}$ denotes the tensor rearrangement operation in $\texttt{einops}$ \cite{rogozhnikov2021einops}, and $l_v$ represents a specific volumetric layer.

In GEM-3D, our models are built upon 2D-to-3D paradigm but treat 3D volume windows as the basic units. The upcoming section will present the process of synthesizing complete volumes, while a more in-depth discussion on the rationale behind the 2D-to-3D paradigm can be found in Appendix \ref{more_comparison}.

\subsection{Informed Slice Conditioned Generation}
\label{sec:3.3}
Unlike cross-modality translation of brain MRI, where the mapping function is relatively constrained \cite{zhu2023make}, reconstructing 3D medical images from mask volumes involves a one-to-many mapping, which can lead to volumetric inconsistency and inferior generation fidelity. To alleviate these issues and improve generation quality, we decouple additional information from the 3D images in the informed slice, which indicates patient anatomical appearance, physical position, and other scanning patterns. During training, the informed slices can be easily drawn from slices in the volume window, while during inference, they are initially generated with models or randomly chosen among accessible volumes and then autoregressively assigned as synthetic ones. Besides guiding the generation process, informed slice-conditioned generation also enables generative enhancement within the dataset through resampling. As a result, condition $\bm c$ in the diffusion model consists of the mask slice $\bm M_{i,j}$ and the informed slice $\bm I_{i,j}$ for the $j$-th slice of the $i$-th volume.

Due to memory constraints, it is impractical to feed the entire volumes into the model. Instead, we randomly sample volume windows from the original volumes. This design further aligns with the idea of introducing variations in Section~\ref{sec:3.4}, as it effectively creates new samples with different conditions in training and introduces variations in starting position for sampling. In detail, for each batch of data, we feed the model the volume window of the $i$-th volume $\bm V_{i, j:j+n} = \{\bm V_{i,j}, \bm V_{i,j+1}, ..., \bm V_{i, j+n-1}\}$, where $\bm V_{i, j:j+n}$ denotes the volume window of the $i$-th sample, starting from the $j$-th slice with a window length of $n$. For each training unit, the informed slice is randomly chosen from $\{\bm V_{i,j}, \bm V_{i,j+1}, ..., \bm V_{i, j+n-1}\}$, and repeated for $n$ times to get $\bm I_{i, j:j+n}$ corresponding to $\bm V_{i, j:j+n}$. 

For denoising the volume window $\bm V_{i, j:j+n}$, we inject the informed slices $\bm I_{i, j:j+n}$ and mask volume $\bm M_{i, j:j+n} = \{\bm M_{i,j}, \bm M_{i,j+1}, ..., \bm M_{i, j+n-1}\}$. In the autoregressive procedure for entire volume sampling, the assigned informed slices always chosen at the endpoints (\ie, $j$ or $j+n-1$) within one window, and the positional relation is ensured via diffusion inpainting, discussed in Section \ref{sec:3.4}. The information from the conditions, i.e., informed slices and mask volume, is integrated via concatenation in the latent space. As for control injection, the simplest method is to concatenate the conditional latents with the noisy target latents. We find this option serves as a simple yet effective solution, and other methods via additional branches \cite{zhang2023adding, mou2023t2i,zhao2024uni} are orthogonal to our method. Since we train the diffusion models from scratch, unlike those methods fine-tuning stable diffusion, simple concatenation is a reasonable and practical choice. See Appendix \ref{abl:feature_injection} for ablation study.

\subsection{Variations in Diffusion Sampling}
\label{sec:3.4}

In the inference stage, GEM-3D enables the synthesis of 3D medical images, controlled with informed slices and mask volumes. To introduce variations and enhance diversity, we discuss several designs.

\para{Informed Slice Selection and Generation.} 
During training, we use the informed slice and mask volume windows from the sample volume. Thanks to the generalizability, with naive combination of conditions, we can sample informed slices from different volumes to drive the conditional generation given the mask volumes and introduce information from the volumes of other patients. To increase variation, a simple method to sample informed slices is to draw the initial slice $\bm I_{i,j}$ from randomly chosen volumes at random starting positions, \ie, $i$ is sampled from $\{1,2,...,N\}$ and $j$ is sampled from $\{1,2,...,Z_i\}$, where $N$ denotes the sample number in the dataset and $Z_i$ denotes the slice number of the $i$-th volume. Another strategy is training a position-conditioned diffusion model to synthesize random informed slices at chosen positions. Then, during inference, we build cascaded diffusion models where we first sample the informed slices condition and then sample the volumes. To accomplish this, we normalize the position information of slices in volumes to ensure the physical meaning of the position information is aligned for different samples. To train a position id guided slice generation model, we use the normalized position id $p \in [0,1]$ and embed it with sinusoidal embeddings as cross-attention condition $\bm p=\operatorname{Embedding}(p)$ \cite{rombach2022high, vaswani2017attention}:
\begin{align}
L(\bm \theta_{s}) = \mathbb{E}_{\bm s_0, \bm p, t, \bm \epsilon \sim \mathcal{N}(0, 1)}\left[\left\| \bm \epsilon - \bm \epsilon_{\bm \theta_s}(\sqrt{\bar\alpha_t} \bm s_0 + \sqrt{1-\bar\alpha_t} \bm \epsilon, t, \bm p) \right\|^2\right],
\end{align}
where $\bm s_0$ is the target slice sample. The slice model $\bm \epsilon_{\bm \theta_s}$ is simply a 2D LDM.

\para{Mask Augmentation.} 
Although simply using the original mask volumes and new informed slices already creates new samples during inference, it is effective and straightforward to further combine augmented mask volumes for more variations. A direct application can be synthesizing counterfactual images for medical diagnostic and educational purposes. For example, we can control the size and position of tumors and synthesize medical scans for different patients. We implement 3D augmentations, including flip, translation, rotation, etc., to the entire mask volume $\bm M_i$ to maintain the volumetric consistency of the condition data.

\para{Bi-directional Propagation from Starting Positions.} 
To sample an entire volume $\hat{\bm V}_i$, we begin by randomly selecting a starting position $p$ and synthesizing the volume window $\hat{\bm V}_{i,j:j+n}$ located at $p$, where $j = \lfloor p \times Z_i \rfloor$ holds, where $\lfloor \cdot \rfloor$ denotes the floor function and $Z_i$ is the slice number of $i$-th volume. We initially sample an informed slice at $p$ via generation or resampling from volumes and use the generated one to serve as the new informed slice recursively to maintain consistency within the same volume. We autoregressively complete the conditional generation with bi-directional propagation, where in each direction, we set an overlapped window length and inpaint the new volume window given the synthesized slices. Specifically, we start with $\hat{\bm V}_{i,j:j+n}$ and then synthesize $\hat{\bm V}_{i,j-h+n:j-h+2n}$, which overlaps with $\hat{\bm V}_{i,j-h+n:j+n}$ to serve as the visible parts in inpainting. The motivation behind this design is to indicate the relative positions between the informed slices and the masked slices, thereby encouraging better volumetric consistency. The implementation follows the basic operation in RePaint \cite{lugmayr2022repaint}, where the noisy known targets are filled in the corresponding logits and combined with generated ones using masks in each iteration. In our setting, we need to place the known noisy latents at the leftmost or rightmost logits since we have two directions. We set the overlapped window length $h$ to be 1 and use that slice as the informed slice recursively. Appendix \ref{infer_algo} presents the process of generating 3D medical images in the enhancement procedure.

\section{Experiments}

\subsection{Experimental Setup} 
\label{sec:4.1}

\para{Datasets.} We evaluate our method on two 3D medical images datasets, namely BraTS \cite{antonelli2022medical, menze2014multimodal} for brain MRI and AbdomenCT-1K \cite{ma2021abdomenct} for abdomen CT. For BraTS, we use FLAIR modality; for the latter, we use subtask 1 in fully supervised task. Since these datasets are provided for medical segmentation challenges and in-house testing ground truth is held, we manually split these paired data and 80\% for training and 20\% for testing. The BraTS dataset comprises 387 training samples and 97 testing samples, while the AbdomenCT-1K dataset consists of 288 training samples and 73 testing samples. 

\para{Implementation Details} After data preprocessing on both two datasets, the majority of volumes contain over 100 slices in the Z-axis, which we further resize to $512\times 512$. 
We implement an improved version of Make-A-Volume \cite{zhu2023make} for comparison, which takes unit window of slices as input to save memory and produce higher quality. We include 3D network-based methods for comparison of generation quality, including a 3D version of Pix2Pix \cite{isola2017image} and a conditional version of 3D latent diffusion models adapted from \cite{khader2022medical, pinaya2022brain}. These methods treat 3D samples as the training unit, and due to the heavy training demand required for $512\times512\times Z$ ($Z$ denotes the normalized slice number in one 3D volume), we omit the implementation of raw 3D diffusion, which removes VAE in LDM. We also exclude other 2D-to-3D medical translation methods like \cite{peng2023generating, han2023medgen3d} due to the lack of code and the fact that our method implicitly integrates the key insights. We design two types of comparison. The first type is to evaluate the generation quality only given the training dataset. The second type is to demonstrate the generalizability and to show the prior information can be propagated with our method. We can evaluate quantitatively given the group distribution for the first type and the paired GT for the second type. The models are trained on 8 NVIDIA V100 32G GPUs. We suggest finding more implementation details including preprocessing details in Appendix \ref{implementation}.

\begin{figure*}[!ht]
\centering
\includegraphics[width=1.00\textwidth]{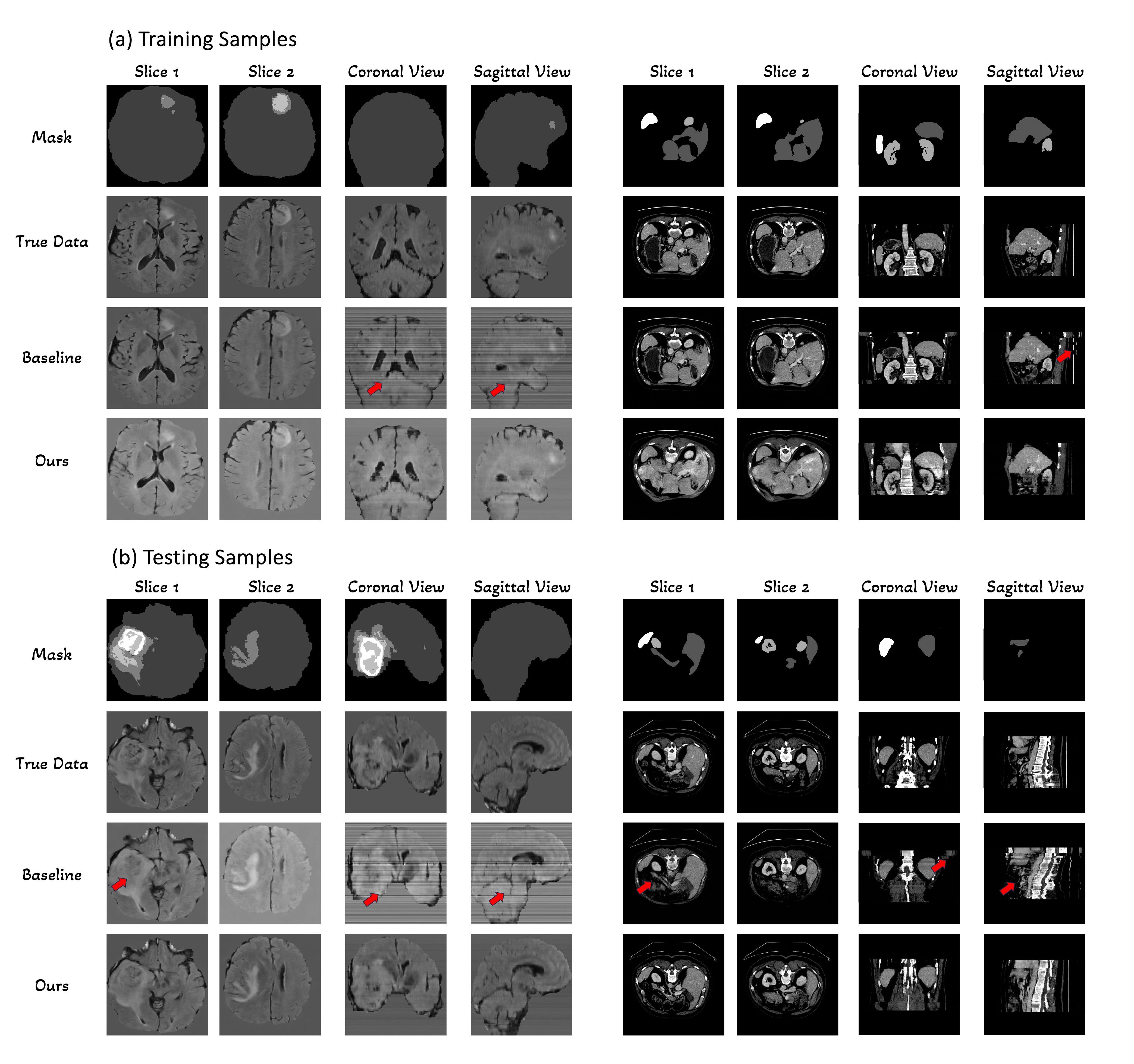}
\caption{\textbf{Qualitative comparison on BraTS and AbdomenCT-1K for training and testing samples.} \textbf{(a)} For training samples, our method synthesizes new samples by introducing variations through randomly chosen informed slices from the given volumes, even when only provided with the training split. In comparison, the baseline method outputs fitting results but still exhibits volumetric inconsistency. \textbf{(b)} For testing samples, our method leverages the additional information of informed slices in the true data and maintains high fidelity, resulting in superior details and improved volumetric consistency comparing with the baseline method.} 
\label{main_comparison}
\vspace{-8pt}
\end{figure*}

\subsection{Qualitative Comparison}

Fig. \ref{main_comparison} provides a qualitative comparison on two datasets for training and testing samples, including the mask for conditions, the true data, the baseline results, and our results. We demonstrate generation fidelity and volumetric consistency of 3D images using two axial slices, coronal view, and sagittal view.
For training samples in Fig. \ref{main_comparison}(a) our method generates novel images with randomly selected slices among the training volumes to serve as the informed slice. GEM-3D enables image generation with variations while maintaining quality and being guided by mask information. In contrast, the baseline method can only output reconstructed images in the training split but fails to achieve realistic visual quality due to volumetric inconsistency (as seen in the coronal and sagittal views). Note that BraTS dataset the color shift is clear among different volumes and it is obvious to observe inconsistency encountered in one-to-many mapping of baseline. However, though the pixel range of AbdomenCT-1K is better normalized, the baseline still produces volumetric inconsistency in the sagittal view indicated by the red arrow. Our method demonstrates higher quality in axial slices and better consistency in 3D by introducing informed slices. Furthermore, we compare results on testing samples in Fig. \ref{main_comparison}(b) to show the informed slice's ability to provide additional patient-prior information. By using the informed slice in the target volume as the initial informed slice in sampling, GEM-3D generates high-quality 3D images that maintain fidelity towards the true data, while the baseline method exhibits poor volumetric consistency and inferior details.

\subsection{Quantitative Comparison}

\begin{table*}[t]
  \centering
  \small
\caption{\textbf{Quantitative comparison on BraTS and AbdomenCT-1K for training samples.} For baseline method, the FID is evaluated on the training samples to show the performance of reconstruction. For our method, we present metrics for different sampling configurations, encompassing initial informed slice selection or generation, and the incorporation of mask augmentation. Darker rows represent sampling results with larger variations compared to the training dataset.} 
  \begin{tabular}{c|ccc|ccc}
    \hline
     & \multicolumn{3}{c|}{BraTS \cite{antonelli2022medical}} & \multicolumn{3}{c}{AbdomenCT-1K \cite{ma2021abdomenct}} \\
    \cline{2-7}

     Methods & FID-A & FID-C & FID-S & FID-A & FID-C & FID-S   \\
    \hline
    Make-A-Volume \cite{zhu2023make}   & 12.96 & 62.63 & 62.04 & 2.66 & 6.68 & 4.25  \\
    \rowcolor[rgb]{ .96,  .96,  .96}  Make-A-Volume + MA  & 15.76 & 71.64 & 73.15 & 2.83 & 7.53  & 5.43 \\
    \hline
    \rowcolor[rgb]{ .92,  .92,  .92}  \textbf{GEM-3D + IG }  & 15.67 & 14.32 & 13.40 & 3.35 & 6.42 & 4.67 \\
    \rowcolor[rgb]{ .88,  .88,  .88}  \textbf{GEM-3D + IG + MA}  & 18.22 & 17.25 & 16.92 & 3.38 & 6.92 & 4.66 \\
    \rowcolor[rgb]{ .94,  .94,  .94}  \textbf{GEM-3D + IC}  &  \textbf{2.07} & \textbf{6.14} & \textbf{6.33} & \textbf{1.81}  & 4.21 & 2.85 \\
    \rowcolor[rgb]{ .90,  .90,  .90}  \textbf{GEM-3D + IC + MA}   & 3.89 & 11.76 & 12.66 & 1.84 & \textbf{4.05} & \textbf{2.69} \\
    \hline
  \end{tabular}
  \label{train_res}
\end{table*}

In Table \ref{train_res}, our results demonstrate a substantial advantage in generation quality. We employ the Frèchet Inception Distance (FID) \cite{heusel2017gans} and calculate the distance between the distributions of generated samples and true samples in the training dataset from three perspectives: axial, coronal, and sagittal, denoted as FID-A, FID-C, and FID-S, respectively. Note that the FID we applied is medical-specific as we use the ResNet-50 \cite{he2016deep} pretrained on the RadImageNet database \cite{mei2022radimagenet}. We generate new data samples using our models under various configurations, including initial informed slice selection or generation, and mask augmentation. MA represents 3D mask augmentation, IG indicates that the initial informed slice is generated, and IC signifies that it is cross-sampled among given volumes. Note that FID comparisons should be conducted within the same dataset. Although the metrics on BraTS are higher than those on AbdomenCT-1K, the generation quality on BraTS slices is superior. 

For the baseline method, we present the reconstruction fidelity on training samples, which can achieve high-quality slice generation performance, making the FID-A of the baseline a good indicator of slice performance. Despite this, FID-C and FID-S, which primarily assess volumetric consistency, reveal that our method exhibits better consistency with the guidance of informed slices.
We also provide optional scenarios for generative enhancement using only the existing dataset, with darker rows representing sampling results with larger variations. Leveraging the slice generation model for initial informed slice introduction results in the most variations without the need for manual informed slice provision. Although this approach leads to a quantitative performance drop due to distribution mismatch for neural models and the existence of error accumulation, FID-A (slightly inferior to baseline) indicates that slice quality is maintained, while FID-C and FID-S show better volumetric consistency. Additionally, through visualization, we observe that the generated informed slices also help produce volumes with competitive quality (see Section~\ref{sec:4.5}).

\begin{table*}[h]
  \centering
  \small
 \caption{\textbf{Quantitative comparison on BraTS and AbdomenCT-1K for testing samples.} For the baselines, we generate 3D medical images from the testing mask volumes. In contrast, for GEM-3D, we further incorporate true informed slice of the testing sample to serve as patient-prior information. D.C. denotes whether the method is able to decouple conditions.}
  \begin{tabular}{c|c|cc|cc}
    \hline
     & & \multicolumn{2}{c|}{BraTS \cite{antonelli2022medical}} & \multicolumn{2}{c}{AbdomenCT-1K\cite{ma2021abdomenct}} \\
    \cline{3-6}
    Methods & D.C. & MS-SSIM $\uparrow$ & LPIPS $\downarrow$ & MS-SSIM $\uparrow$ & LPIPS $\downarrow$  \\
    \hline
    Pix2Pix 3D \cite{isola2017image}  & \ding{55} &  0.776  & 0.277 & 0.634 & 0.375  \\
    3D LDM \cite{khader2022medical, pinaya2022brain}  & \ding{55}  & 0.803  & 0.259 & 0.614 & 0.337  \\
    Make-A-Volume \cite{zhu2023make}  & \ding{55}  & 0.817  & 0.234 & 0.620 & 0.318 \\
    \hline
    \textbf{GEM-3D}  & \ding{51} & \textbf{0.887} & \textbf{0.134} & \textbf{0.727} & \textbf{0.204} \\
    \hline
  \end{tabular}
  \label{test_res}
\end{table*}

To validate the effectiveness of the informed slice approach, which aids in guiding the generation process with patient-prior information. To do this, we conduct a comparison on testing samples. The baselines involve generating 3D medical images from the test mask volumes, while our proposed method combines the true informed slice from the testing volume, which is then assigned as the initial informed slices. 
As the synthesized slices serve as the new informed slices in an autoregressive manner, the patient-wise initial information is propagated throughout the generation process, thereby assisting in guiding the anatomical information. This results in the synthesized images bearing a close resemblance to the ground truth, as illustrated in Fig. \ref{main_comparison}.
To measure the similarity between the ground truth volumes and the generated volumes, we employ two metrics: MS-SSIM \cite{wang2003multiscale} and LPIPS \cite{zhang2018unreasonable}. These metrics are computed based on axial slice pairs. Table \ref{test_res} indicates that the informed slice approach significantly enhances the generation fidelity by incorporating patient-prior information. The comparison between the baselines also ensure the quality of 2D-to-3D foundations and please refer further discuss in Appendix \ref{more_comparison}.

\begin{figure*}[!h]
\centering
\includegraphics[width=0.65\textwidth]{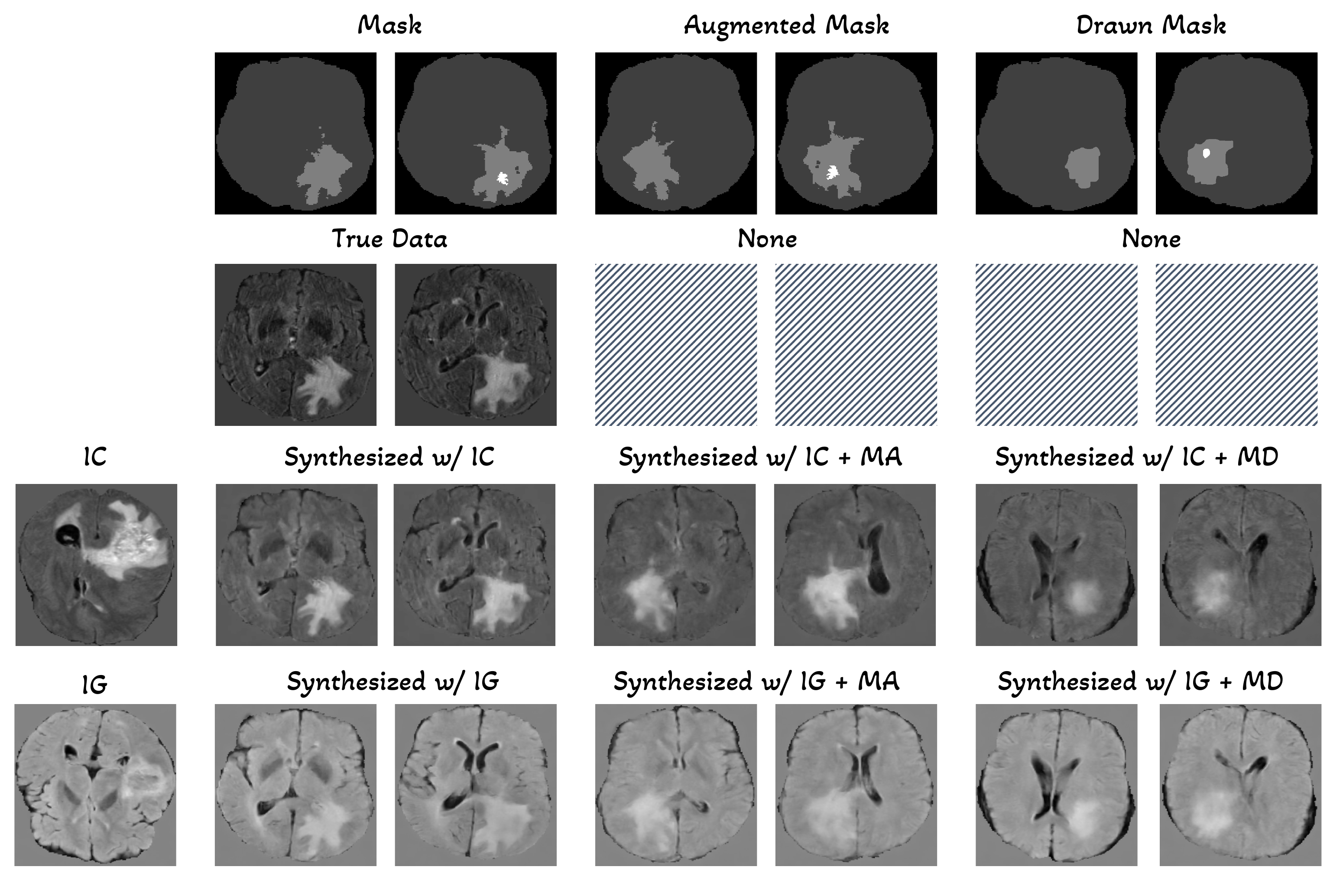}
\caption{\textbf{GEM-3D enables counterfactual synthesis.} We show synthesis results under different scenarios with respect to informed slices and masks. The proposed method produces high-fidelity generation performances and offers solutions for counterfactual synthesis under different occasions.} 
\label{counterfactual}
\vspace{-8pt}
\end{figure*}

\begin{figure*}[!h]
\centering
\includegraphics[width=0.7\textwidth]{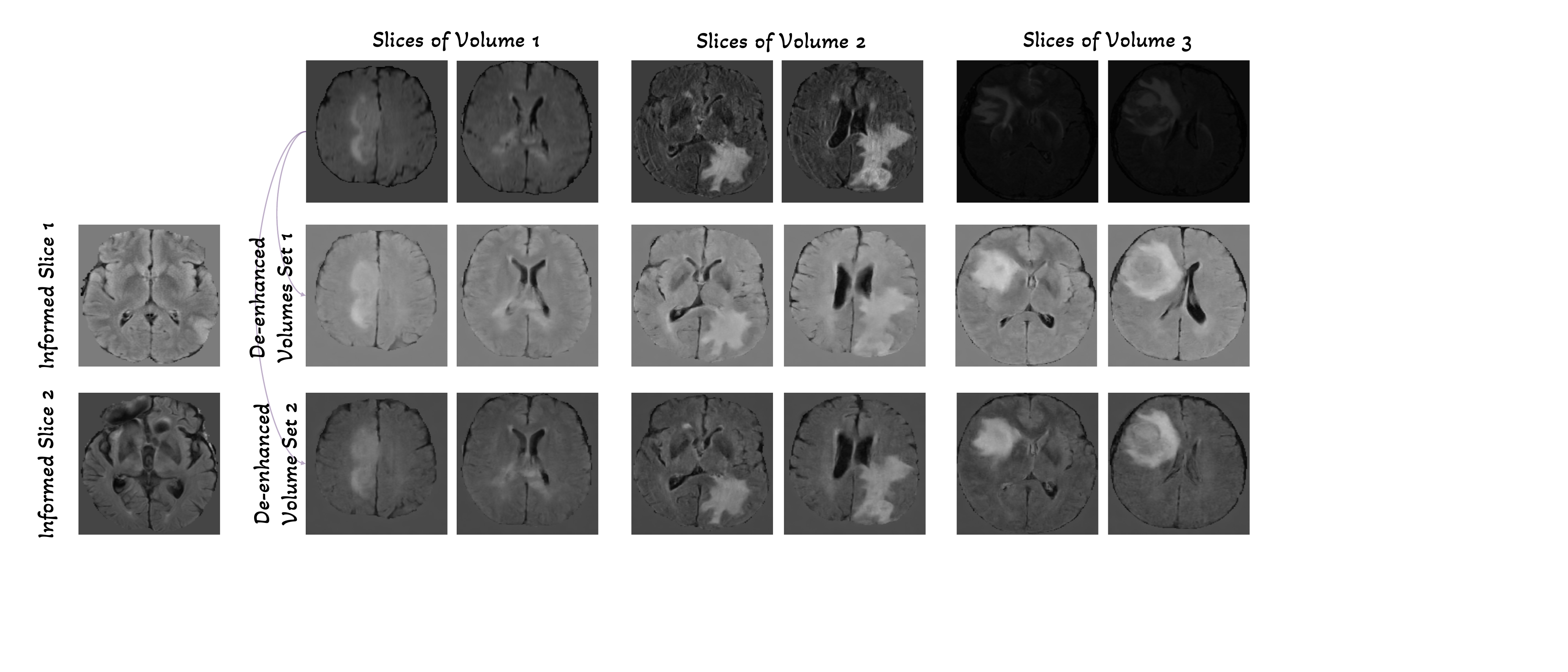}
\caption{\textbf{GEM-3D enables generative de-enhancement.} We show three volumes samples on BraTS with the last sample exhibiting poorer visualization quality due to the absence of normalization. Using two choices of informed slices for the initial sampling, GEM-3D generates two types of well-normalized entire volumes through de-enhancement. It is important to note that counterfactual samples are generated for dataset-level normalization, and thus, they are not required to maintain all the anatomical details as observed in the true samples (first row).} 
\label{de-enhance}
\vspace{-8pt}
\end{figure*}

\subsection{Further Evaluation}
\label{sec:4.5}

\para{GEM-3D Enables Counterfactual Image Synthesis.} 
By decoupling informed slices and mask volumes, GEM-3D introduces control over masks and facilitates synthesis with counterfactual masks. As shown in Fig. \ref{counterfactual}, we generate counterfactual masks using 3D augmentations and manual drawing, and employ either cross-selected or generated informed slices. Despite slightly inferior metrics for informed slices combined with generation compared to cross-selection, the slice diffusion model produces high-quality informed slices, and the cascaded model generates well-synthesized results from qualitative demonstrations. Remarkably, even with manually drawn masks, the synthesized results maintain high fidelity, suggesting potential for doctor-assisted control. Our method effectively handles various scenarios, highlighting its potential for future applications.

\para{GEM-3D Enables De-enhancement.} 
Moreover, we explore the concept of de-enhancement as a normalization process, particularly relevant in medical datasets. Often, these datasets necessitate post-hoc registration and normalization due to variations in imaging protocols, scanner hardware, and patient positioning that can result in inconsistent intensity distributions. We demonstrate GEM-3D can effectively de-enhance the dataset using informed slice control, achieving medical harmonization in the dataset level. This technique is similar to other methods \cite{liu2021style} that employ GANs to transfer the style of MRI images.
To showcase the efficacy of GEM-3D in generating well-normalized volumes, we synthesize volumes with the same informed slice, as illustrated in Fig. \ref{de-enhance}. This figure displays three volume samples from the BraTS dataset, wherein the last sample exhibits poorer visualization quality due to the lack of normalization. GEM-3D is capable of generating two types of well-normalized volumes through de-enhancement with two choices of informed slices. 
This de-enhancement approach offers significant contributions to data processing in medical domains, facilitating applications in MRI harmonization for multi-site hospitals. 

\section{Conclusions}

In this paper, we present GEM-3D, a novel approach for synthesizing high-quality, volumetrically consistent 3D medical images and enhancing medical datasets with diverse diffusion sampling variations. Our key insight is the design of informed slices, which effectively guide volumetric sampling and enable decoupled controls. Consequently, GEM-3D generates realistic 3D images from existing datasets and offers solutions for counterfactual synthesis with mask controls.

\newpage
\appendix
\appendix

\noindent \textbf{Roadmap.} In Appendix, we present a comprehensive analysis and additional results of GEM-3D. Section \ref{implementation} provides more implementation details, and Section \ref{infer_algo} describes the inference algorithm. We also conduct extensive ablation studies in Section \ref{abl} to assess the effectiveness of overlapped inpainting, volumetric tuning, and feature injection. An in-depth analysis on 2D-to-3D foundations of our method is presented in Section \ref{more_comparison}, along with a downstream task in Section \ref{more_seg}. Furthermore, we provide more qualitative results and limitations and future work in Section \ref{more_res}. Finally, we address the ethical considerations pertinent to our research in Section \ref{ethic}.

\section{Implementation Details}
\label{implementation}
We utilize the nnU-Net library \cite{isensee2021nnu} for preprocessing the data, including generating non-zero masks and resampling the 3D images. The non-zero masks, corresponding to the brain contour in brain MRI, are used to ensured that all mask slices in the BraTS dataset are non-empty. 
Both the baseline method and ours have slice models trained for 150k iterations, and volumetric layers tuned for 50k iterations, with the AdamW optimizer \cite{loshchilov2017decoupled}. The overall training costs one week. The window length of volume windows (\ie, the training samples) is 16. 
The optional 3D augmentations comprise random flip, rotation, and translation applied to the tumor for BraTS. The rotation is limited to ±2.5 degrees on each axis, while translation moves the tumor towards the 3D center of the volume, uniformly sampling within a 60-pixel range. To prevent meaningless masks, abdominal CT data only has 3D rotation.
For sampling, we adopt DDIM \cite{song2020denoising} with 200 steps.

\section{Inference Algorithm}
\label{infer_algo}

Algorithm \ref{alg:sampling} shows the overall procedure for diffusion sampling in GEM-3D. With the trained model, GEM-3D allows initiating at a randomly selected position $p$ and propagating bi-directional sampling through the operations of the RePaint algorithm \cite{lugmayr2022repaint} on overlapping slices. The primary motivation for integrating RePaint is to enhance the consistency of the 3D volume and enable the patient's prior information in the informed slice to propagate in a 2D-to-3D manner.

\section{Ablation Studies}
\label{abl}
\subsection{Ablation on Overlapped Inpainting}
In inference, we sample a window of slices and use bi-directional propagation to form the entire volumes. In this process, we set a overlapped window and repaint the overlapped slices via RePaint \cite{lugmayr2022repaint}. This approach effectively enhances volumetric consistency within the resulting volumes.

\begin{figure*}[!h]
\centering
\includegraphics[width=1.0\textwidth]{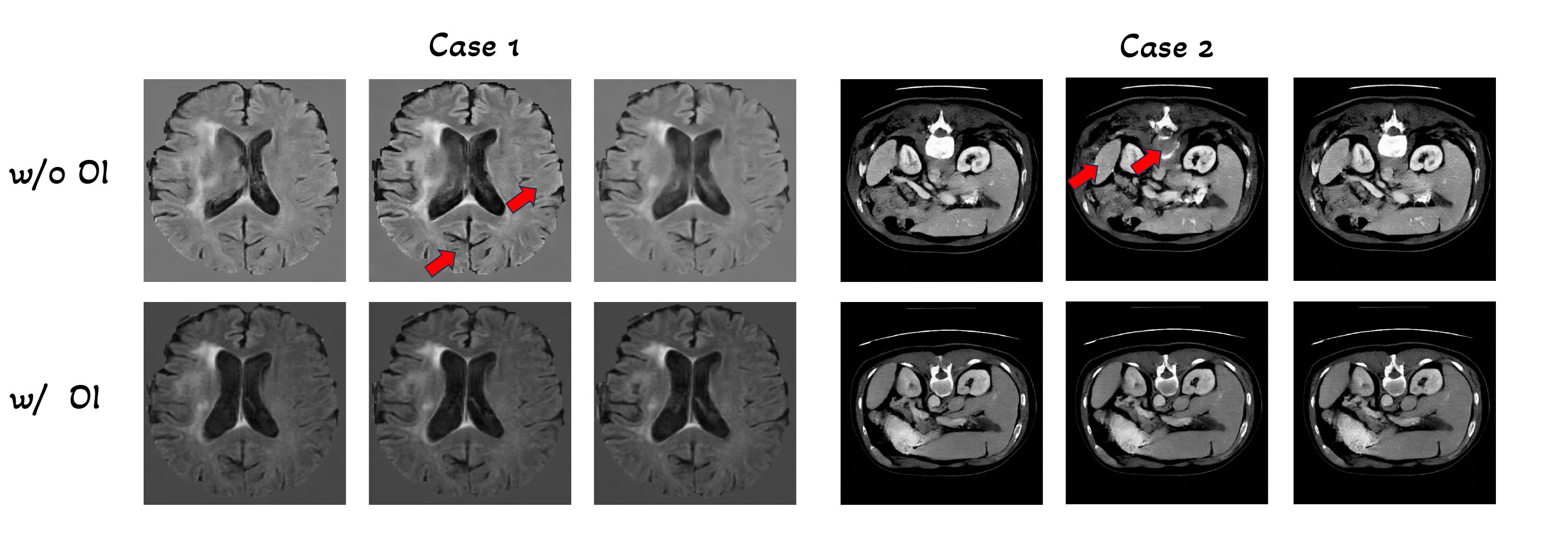}
\caption{\textbf{Ablation on Overlapped Inpainting.} We present two cases for comparison, with each case displaying three consecutive slices in a sample. OI refers to overlapped inpainting.} 
\label{oi}
\vspace{-8pt}
\end{figure*}

In training, we feed the model a window of slices and a randomly selected informed slice within the window. As a result, the relative position between the informed slice and the data can be random within a certain range. For sampling, since the informed slice is located at the overlapping position except for the first sampling, we can control the relative position to enhance the volumetric performance through inpainting in the autoregressive process. Fig. \ref{oi} shows the ablation study on overlapped inpainting.
We observe that in a volume, some consecutive slices exhibit sharp visual effects or inconsistent shape changes, indicating inconsistency within the volume. This inconsistency is apparent when using visualization software like ITK-SNAP \cite{py06nimg}, and it occurs due to the inconsistency between each sampling within a single volume. On the contrary, by autoregressively injecting the relative position information, we can achieve better consistency through overlapped inpainting.

\begin{algorithm}[t]
    \algsetup{linenosize=\small}
    \caption{Code for Diffusion Sampling}
    \label{code}
    \begin{algorithmic}[1]
    \label{alg:sampling} 
    \STATE \textbf{Require} volume diffusion model $\bm \epsilon_{\bm \theta_v}$, slice diffusion $\bm \epsilon_{\bm \theta_s}$, pre-trained VAE $\mathcal{E}$ and $\mathcal{D}$
    \STATE \textbf{Input} mask volume $\bm M_i$, and the starting position $p$ for sampling 
    \STATE \small{\color{gray}// First Sampling}
    \STATE Given $p$, sample informed slice $\bm I$ among $\{\bm V_i\}_{i=1}^{N}$ and get $\bm c_{\bm I} = \mathcal{E}(\bm I)$
    \STATE \textbf{or} directly set $\bm c_{\bm I} = \hat{\bm s}_0$ with $\bm \epsilon_{\bm \theta_s}$

    \STATE $\hat{\bm M_i} = \operatorname{Augmentation3D}(\bm M_i)$
    
    \STATE Initialize noises $\{\bm x_j, \bm x_{j+1}, \ldots, \bm x_{j+n-1}\}_T \sim \mathcal{N}(\bm{0}, \bm{I})$ corresponding to $\hat{\bm V}_{i,j:j+n}$ \\
    \STATE Set $\bm c_{j+k} = \operatorname{Concat}(\mathcal{E}(\hat{\bm M}_{i,j+k}), \bm c_{\bm I})$ for $k$ in  \texttt{Range}(n)\\
    \STATE Denoise to $\{\bm x_j, \bm x_{j+1}, \ldots, \bm x_{j+n-1}\}_0$ with $\bm \epsilon_{\bm \theta_v}$ conditional on $\bm c_j, \bm c_{j+1}, \ldots, \bm c_{j+n-1}$ \\

    \STATE \small{\color{gray}// Bi-directional Propagation}
    \FOR{$dir$ in $\{\operatorname{UP}$, $\operatorname{DOWN}\}$}
        \STATE \small{\color{gray}// $\operatorname{NUM}$ is determined by $p$, $n$, $dir$, and slice number $Z_i$} \\
        \FOR{$iter = \texttt{Range}(\operatorname{NUM}(p, n, dir, Z_i))$}

        \STATE Save $\{\bm x_{j^\prime}\}_0$ \textbf{or} $\{\bm x_{j^\prime+n-1}\}_0$ known in the last sampling \\
        \STATE Initialize noises $\{\bm x_{j^\prime}, \bm x_{j^\prime+1}, \ldots, \bm x_{j^\prime+n-1}\}_T \sim \mathcal{N}(\bm{0}, \bm{I})$ 
        \STATE Denote $\bm w_T = \{\bm x_{j^\prime}, \bm x_{j^\prime+1}, \ldots, \bm x_{j^\prime+n-1}\}_T $ \\
        \STATE Denote $\bm o_0 = \{\bm x_{j^\prime}\}_0$ \textbf{if} ($dir$ is $\operatorname{UP}$) \textbf{else} $\{\bm x_{j^\prime+n-1}\}_0$ \\ 
        
        \STATE Set $\bm c_{\bm I} = \{\bm x_{j^\prime}\}_0$ \textbf{if} ($dir$ is $\operatorname{UP}$) \textbf{else} $\{\bm x_{j^\prime+n-1}\}_0$ \\ 
        \STATE Set $\bm c_{j^\prime+k} = \operatorname{Concat}(\mathcal{E}(\hat{\bm M}_{i,j^\prime+k}), \bm c_{\bm I})$ for $k$ in \texttt{Range}(n) \\

            \FOR{$t = T, \ldots, 1$}
                \STATE $\bm \epsilon, \bm z \sim \mathcal{N}(\bm{0}, \bm{I})$ \textbf{if} $t > 1$ \textbf{else} $\bm \epsilon, \bm z = \bm{0}$ \\
                \STATE $\bm o_{t-1} = \sqrt{\bar\alpha_{t-1}} \bm o_0 + \sqrt{1-\bar\alpha_{t-1}} \bm \epsilon $ \\
                \STATE $\bm w_{t-1} = \frac{1}{\sqrt{\alpha_t}}\left( \bm w_t - \frac{1-\alpha_t}{\sqrt{1-\bar\alpha_t}} \bm{\epsilon}_{\bm \theta_v}(\bm w_t, t, \bm c) \right) + \sigma_t \bm{z}$ \\
                \STATE $\bm w_{t-1}[0] = \bm o_{t-1}$ \textbf{if} $dir$ is $\operatorname{UP}$ \textbf{else} $\bm w_{t-1}[-1] = \bm o_{t-1}$ \\
                
            \ENDFOR
        \ENDFOR \\
    \ENDFOR \\
    \STATE \small{\color{gray}// Decoding}
    \STATE Obtain $\{\bm x_0, \bm x_1, \ldots, \bm x_{Z_i-1}\}_0$ \\
    \STATE Return $\hat{\bm V}_{i} = \mathcal{D}(\{\bm x_0, \bm x_1, \ldots, \bm x_{Z_i-1}\}_0)$
    
    \end{algorithmic}
\end{algorithm}

\subsection{Ablation on Volumetric Tuning}

We build our method upon volume diffusion, which employs a two-stage tuning process and utilizes volumetric layers to enhance volumetric consistency in 3D medical images \cite{zhu2023make}. While this aspect is not the main contribution of our work, we include the ablation study here.

\begin{figure*}[!h]
\centering
\includegraphics[width=1.0\textwidth]{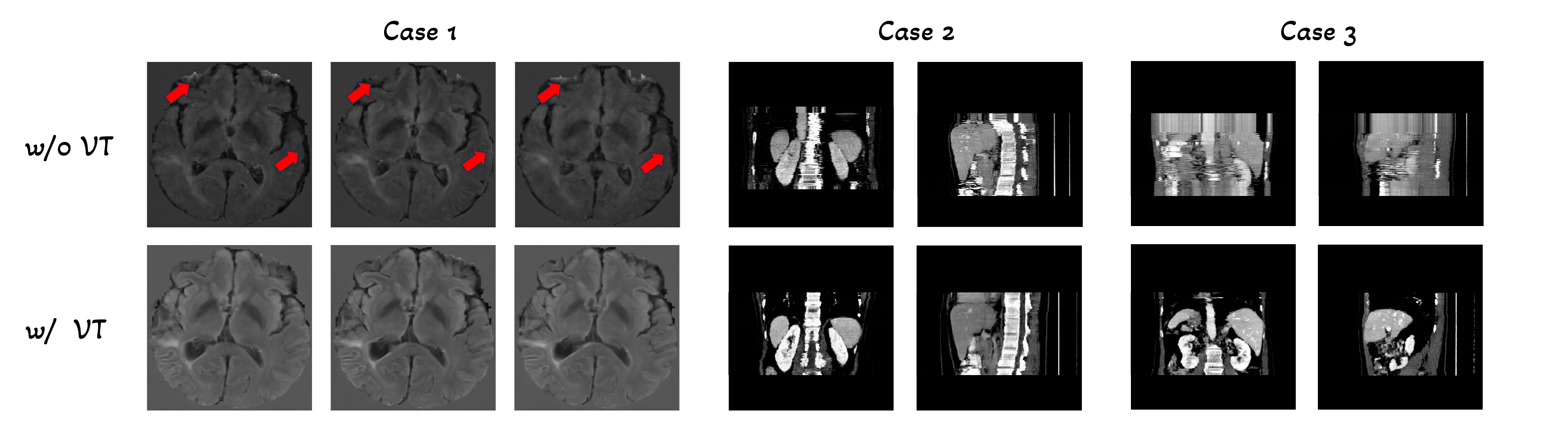}
\caption{\textbf{Ablation on Volumetric Tuning.} We present three cases for comparison. In case 1, we display three consecutive slices from a sample in the BraTS dataset. For cases 2 and 3, we exhibit the coronal and sagittal views from the AbdomenCT-1K dataset. VT denotes volumetric tuning.} 
\label{vt}
\vspace{-8pt}
\end{figure*}

Fig. \ref{vt} demonstrates the effectiveness of volumetric tuning in mitigating volume inconsistencies. In case 1, red arrows highlight the inconsistency among three consecutive slices. For cases 2 and 3, coronal and sagittal views are presented to clearly show the improvement in consistency. Notably, in case 3, without volumetric tuning, a poor-quality failure case is produced.

\subsection{Ablation on Feature Injection}
\label{abl:feature_injection}

Although the method for incorporating condition features is orthogonal to our approach, we perform ablation studies on various feature injection techniques. We have two types of spatial conditions, namely, the mask and the informed slice. To compare these conditions, we examine simple concatenation and integration of control branches (e.g., ControlNet \cite{zhang2023adding}) as well as cross-attention with CLIP \cite{radford2021learning}. When implementing ControlNet for two spatial information types, we diverge from the original approach of freezing the main branch and tuning the adapter; instead, we adjust both the main and condition branches simultaneously due to the absence of foundation backbones like Stable Diffusion \cite{rombach2022high}. In the case of combining input-level concatenation and CLIP, we opt for the more technically sound approach of concatenating mask conditions and embedding the informed slice with cross-attention, as mask conditions provide a stronger spatial correspondence with the data.

Table \ref{feature_injection} demonstrates that all the different methods yield comparable performance. It is reasonable that ControlNet, a powerful technique for adding new conditions to a given diffusion model, does not produce significantly higher results, considering the distinct training strategy necessitated by the absence of a pretrained foundation model.

\begin{table*}[h]
  \centering
  \small
 \caption{\textbf{Quantitative comparison on BraTS for testing samples.} }
  \begin{tabular}{c|cc}
    \hline
    Methods & MS-SSIM $\uparrow$ & LPIPS $\downarrow$ \\
    \hline
    ControlNet \cite{zhang2023adding}  & 0.844  & 0.140\\
    Concatenation + CLIP \cite{radford2021learning} &  0.879  & 0.135  \\
    Concatenation  & \textbf{0.887} & \textbf{0.134} \\
    \hline
  \end{tabular}
  \label{feature_injection}
\end{table*}

\section{More Analysis on Comparison Methods}
\label{more_comparison}

As demonstrated in Table \ref{test_res}, baseline approaches yield inferior results in terms of slice quality. Although GAN-based methods have achieved great success in recent years, they often suffer from unstable training and mode collapse. For initial attempts with diffusion models in medical generation, it is challenging to adopt 3D medical volumes as training samples due to the high memory demand. In contrast, our method employs efficient diffusion models with a pseudo 3D architecture and uses volume windows as the training unit, enabling model training with nearly 30G GPU memory. Notably, these methods cannot decouple conditions, including patient prior and mask volumes, preventing the generation of new 3D volumes using only training datasets.

Besides lower training demand, we emphasize other reasons for adopting a 2D-to-3D approach in our method. Firstly, the rationale is closely related to our information propagation, which initially relies on an informed slice to indicate patient prior and then propagates the information across 3D volumes. Secondly, we believe that the 2D-to-3D approach does not lead to performance degradation, as existing methods on limited datasets have already demonstrated superior performance compared to 3D-based methods, despite the presence of some special patterns (strip artifacts) in failure cases. Progress in video generation has shown that the performance of pseudo-3D backbones can be impressive, given the technical foundations of video generation \cite{singer2022make, blattmann2023align, girdhar2023emu}. Lastly, during inference, our method permits starting at a random position and propagating the sampling, making it highly compatible with practical scenarios. In real-world situations, medical scanning is often unstructured, capturing different zones for different patients and resulting in varying volume start and end scanning positions, particularly when involving scanners from multiple hospitals. Such cases necessitate post-hoc registration and normalization for 3D medical images. Our proposed method addresses these challenges and provides better alignment with realistic scenarios.

\section{More Evaluation}
\label{more_seg}

\para{GEM-3D Enables Segmentation with Reduced Privacy Risks.} By re-sampling the original dataset, we can generate a new version of the segmentation dataset with reduced privacy risks for downstream task training. To address privacy concerns, our method relies solely on generated data while preserving the 3D segmentation masks. Specifically, we employ distinct informed slices and use the true mask volumes to guide the re-sampling process, resulting in the desensitization of medical data. Consequently, all generated volumes maintain the physical relationship between mask and data but remove patient-specific details.

We assess our method's segmentation performance using nnU-Net \cite{isensee2021nnu} on the BraTS dataset and compute the mean HD95 and Dice scores for WT, ET, and TC, as defined in \cite{hatamizadeh2022unetr}. We implement data resampling up to three times and treat all volumes as a training dataset. As a result, there are three times training samples than the True Data, but the mask volumes remain the same with different generated data. This setting corresponds to (3x) in Table \ref{seg}, and we also report the results of (1x) and (2x) using datasets generated similarly. For comparison, we apply the reconstructed data from the baseline method (\ie, Make-A-Volume), which has inferior quality compared to True Data and cannot desensitize the data. Also note there is no need to sample multiple times for the baseline method because only the reconstructed data is given. As shown in Table \ref{seg}, our method yields significantly better results than the baseline reconstruction data, which still exhibits volumetric inconsistencies. Furthermore, our method demonstrates a moderate performance drop (-1.33\% Dice and +0.17mm HD95) compared to using the actual training data, striking a balance between segmentation accuracy and privacy preservation. Notably, the results with (1x) and (2x) exhibit a higher performance drop since they cannot achieve better training distribution coverage than (3x), but they benefit from increased inference efficiency.

\begin{table*}[t]
  \centering
  \small
 \caption{\textbf{Quantitative comparison on segmentation performances of BraTS.} We report Dice and HD95 on test split of BraTS using nnUNet.}
  \begin{tabular}{c|cc|cc}
    \hline
    Training Data & Dice (\%) $\uparrow$ & $\Delta$ Dice (\%) $\uparrow$ & HD95 (mm) $\downarrow$ & $\Delta$ HD95 (mm)$\downarrow$  \\
    \hline
    True Data  & 72.90  & / & 8.89 & / \\
    \hline
    Reconstructed Data of Baseline & 59.70 & -13.20 & 12.44 & 3.55 \\
    Re-sampled Data of GEM-3D (1x) & 68.53 & -4.37 & 10.04 & 1.15 \\
    Re-sampled Data of GEM-3D (2x) & 71.13 & -1.77 & 9.35 & 0.46 \\
    \textbf{Re-sampled Data of GEM-3D (3x)} & \textbf{71.57} & \textbf{-1.33} & \textbf{9.06} & \textbf{0.17} \\
    \hline
  \end{tabular}
  \label{seg}
\end{table*}

\section{More Results and Discussion}
\label{more_res}

\subsection{Synthetic Results}
\begin{figure*}[!h]
\centering
\includegraphics[width=1.0\textwidth]{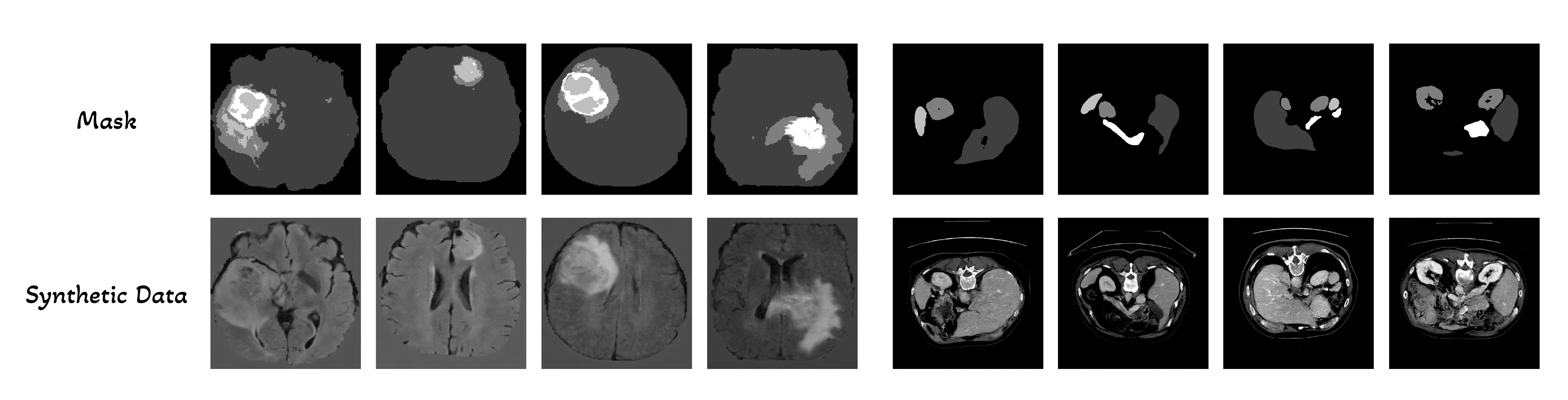}
\caption{\textbf{More Results of Synthetic Data.} We show slice results of the synthetic data and the corresponding mask slice.} 
\label{more}
\vspace{-8pt}
\end{figure*}

Fig. \ref{more} displays paired new samples from two datasets. As some samples lack true data for reference (augmented masks as conditions), only masks and synthetic slices are shown. Although slice results do not fully represent 3D results, especially when multiple classes are present in the masks, the high-quality slices correspond well with the masks.

\subsection{Failure Cases}
\begin{figure*}[!h]
\centering
\includegraphics[width=0.7\textwidth]{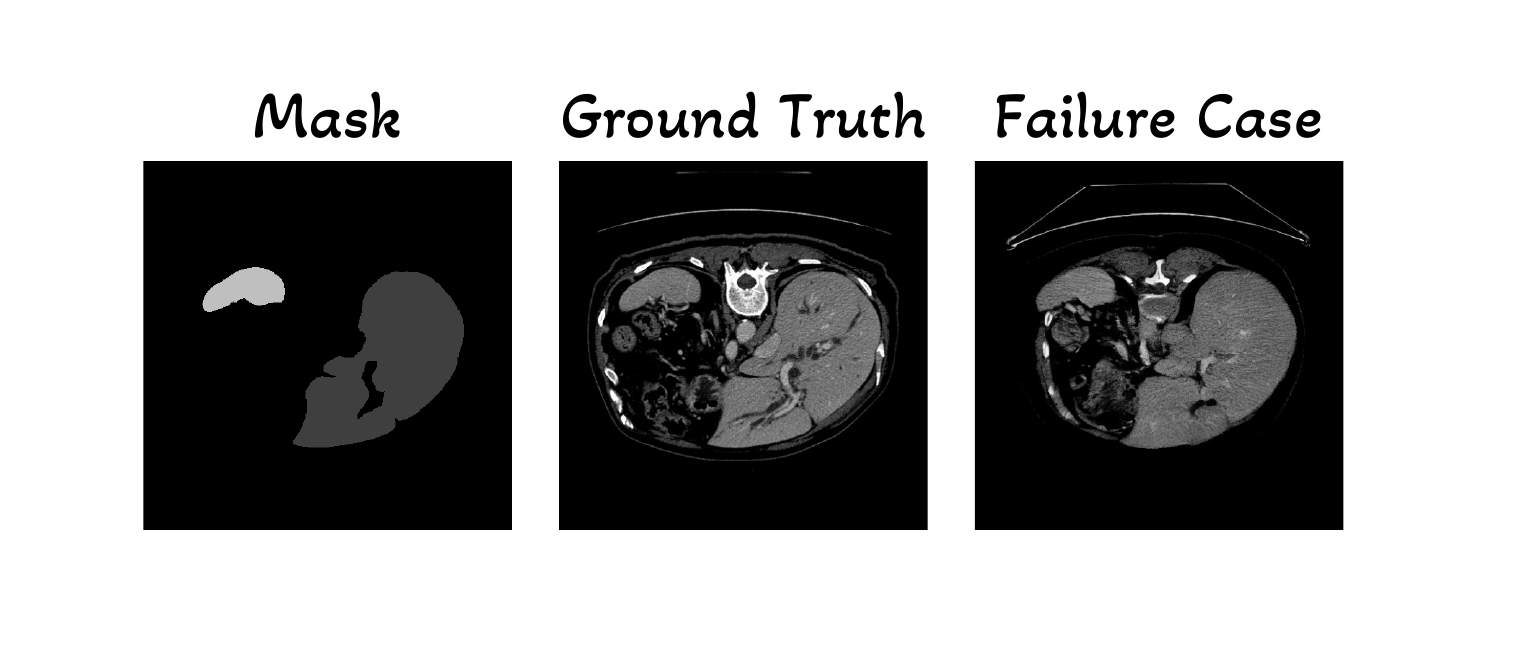}
\caption{\textbf{Failure Results on AbdomenCT-1K.} In the failure results on the AbdomenCT-1K dataset, the mask, true data, and synthetic data are displayed. The failure case appears unrealistic and does not adhere to anatomical rules.} 
\label{failure_fig}
\vspace{-8pt}
\end{figure*}

We have observed a typical type of failure cases on AbdomenCT-1K, as illustrated in Fig. \ref{failure_fig}. In these cases, our method generates organs that tend to be located at the edge of the human body or even intersect with each other, violating anatomical rules and producing unrealistic samples. We believe this issue arises due to a significant mismatch between the initial informed slice and the organ mask. 
The informed slice carries information about the human body shape through learning (which is somewhat determined by the training dataset), while the organ masks indicate the precise organ locations. When an informed slice corresponds to a disproportionately small shape, it may result in this type of failure cases. To address this issue, post-processing techniques can be applied to limit the degree of mismatch between the informed slice and the organ masks, or alternatively, the informed slices and masks can be automatically edited.

\subsection{Limitations and Future Work}
\label{limitations}
The performance of diffusion models is limited by the size of 3D medical image datasets, restricting their potential for broader applications. In our experiments, we use hundreds of volumes for training, yielding impressive in-domain inference results. However, larger datasets could further enhance the models' generalization capabilities across different medical modalities, organs, and other factors. We believe that developing advanced foundational models for 3D medical images is crucial and beneficial for the research community. Additionally, it is essential to combine generation models with more downstream tasks to boost higher performances than the state-of-the-art results when only using synthetic data, addressing privacy concerns in the medical domain.

\section{Ethic Issues}
\label{ethic}

\subsection{Dataset}
\label{ethic_data}
We utilize existing datasets, including a combined version of BraTS \cite{menze2014multimodal} from the Medical Segmentation Decathlon \cite{antonelli2022medical} and AbdomenCT-1K \cite{ma2021abdomenct}. These datasets involve human patients, as they consist of scans of human brains and abdomens obtained through CT or MRI imaging. To the best of our knowledge, the collection of these datasets and the use of the data are legal and have been conducted under appropriate supervision.

\subsection{Broader Impacts}
\label{broader}
We present a generative method for synthesizing CT or MRI scans of specific patients, offering a practical solution to enhance medical datasets, an under-explored yet critical area. Additionally, we demonstrate competitive downstream results in Appendix \ref{more_seg}, where using only generated data, we can achieve comparable segmentation performances with the same method, thereby mitigating privacy risks in medical literature. However, this approach also raises ethical considerations. While this does not directly related to fraud, it can also be treated as a kind of deepfake and raises important ethical considerations. The creation of realistic medical images could lead to concerns about privacy and consent, as individuals may not be aware that their personal health information is being used to generate synthetic data. Additionally, the misuse of such synthetic data could have serious consequences, such as the attack to hospital data system that affects the practices in healthcare. It is crucial to address possible ethical issues and establish guidelines for the responsible use of generative methods in medical imaging to ensure that the benefits of this technology are realized without causing harm to individuals or society.

{
\small
\bibliographystyle{unsrt}
\bibliography{neurips_2024}
}

\end{document}